# Experimental and computational investigation of single particle behavior in low Reynolds number linear shear flows

Nima Fathi[1*], Seyed Sobhan Aleyasin[2], Peter Vorobieff[1], and Goodarz Ahmadi[3]

[1] *Department of Mechanical Engineering, University of New Mexico, Albuquerque, NM, USA*
[2] *Department of Mechanical Engineering, University of Manitoba, Winnipeg, MB, Canada*
[3] *Mechanical & Aeronautical Engineering, Clarkson University, Potsdam, NY, USA*

*Corresponding author email: nfathi@unm.edu

**Abstract**

Trajectories of a buoyant spherical solid particle in a linear shear flow were investigated at low Reynolds numbers. A two-dimensional CFD analysis was performed to simulate the solid-fluid flows. Our numerical model, the discrete phase element method, was used to model and simulate the fluid domain and particle motion as the solid phase. The reliability of the computational results was evaluated for the particle trajectory. The agreement between the numerical results with the experimental data was quantified.

*Keywords:* particle migration, multiphase flows, linear shear flows, CFD

## 1. Introduction

A fluid-solid flow is an interdisciplinary research area with many technological, commercial and medical applications. Suspensions of macro- to nano-scale particles in viscous fluid flows occur in transport of sediments in rivers and estuaries, to the use of nano-fluids as high-performance coolants. Other important applications are contaminant transport and exposure assessment and slurries flows, as well as, secondary oil recovery by hydraulic fracturing. The behavior of solid particles in a viscous fluid is one of the oldest classical problems in this field. One of the earliest investigations of the motion of small particles in a viscous fluid at low Reynolds number ($Re$) was performed by Stokes in 1851 [1]. More recently, Ingber et al. [2] investigated a variety of particle interactions including particle/particle, wall/particle and particle migration in nonlinear shear fields. They also developed a semi-analytical solution for the motion of two spherical particles suspended in an unbounded arbitrary shear flow [3]. Comparisons of computational prediction of single and double particle trajectories using the discrete phase model (DPM) against the semi-analytical solution were performed in [4,5,6]. In addition to DPM, several particle methods including smoothed particle hydrodynamics, smoothed profile hydrodynamics and the modified version of the front tracking method have been utilized to investigate the migration of macro- to micro-scale droplets, bubbles and solid particles in fluid flows [7-11]. In these numerical approaches, accurate evaluation of the discontinuities between the primary and the secondary phases can be challenging.

In this article, our new development of the linear shear flow in the Couette flow apparatus is presented. The modified arrangement makes it possible to study the single and multi-particles migration in parallel and counter Couette flows. The new device also provides higher accuracy for setting the boundary conditions for making fluid and particle measurement compared to previous facilities of the team.

## 2. Experimental and computational procedure

To experimentally determine the behavior of a single spherical particle suspended in linear shear flow, one spherical polymethyl-metacrylate (PMMA) particle with a diameter of 6.35*mm* was suspended into a tank between two layers of fluids with approximately equal viscosity but with different densities. The particle was placed between two belts as shown in the top view of the physical domain in Fig. 1. The schematic representation of the experiment showing the boundary conditions of the physical domain and the suspended particle between the two layers is also shown in Fig. 1. After the particle is settled in the desired position, the side belts are moved at a constant velocity to provide the fluid domain with a shear field in the Couette flow configuration. A computer-controlled stepper motor is used to provide the moving boundary conditions (moving belts). The rectangular tank is filled with a stratified viscous fluid which is a water solution of $ZnCl_2$ and Triton X100. The fluid properties of the primary phase – upper and lower layer – are shown in Table 1. The fluid components used in this analysis were examined and evaluated numerous times by the team. The same stratified fluid components were also utilized to study the migration of pairs of nearly spherical solid particles suspended in a shear flow inside a Couette cell for the Reynolds number of approximately 0.1 [12, 13].

To model and simulate the fluid domain including the solid particle phase, the standard discrete phase model is used. The Lagrangian DPM is based on a translational force balance that is formulated for an individual particle. In the DPM, typically the particle is subject to gravity, drag, pressure, Magnus, virtual mass and Saffman forces.

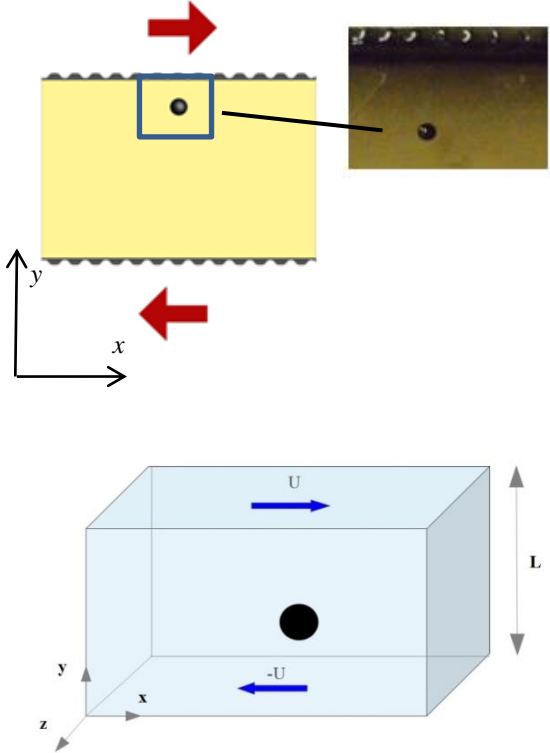

**Fig. 1** Top: top view of the single particle and the upper moving belt; bottom: oblique view of schematic fluid and solid particle domain

**Table 1** Stratified fluid properties

|  | $\rho\ (kg/m^3)$ | $\nu\ (cm^2/sec)$ |
|---|---|---|
| Upper layer fluid | 1110 | 146.2 |
| Lower layer fluid | 1280 | 146.2 |

In this method, the influence of particle is represented as a source term in the Navier-Stokes equations. The continuous phase equations are solved in conjunction with the tracking of particles. For dilute systems, the effect of particles on the flow is negligible and the source term is set to zero. The continuous phase equations are,

$$\frac{\partial \rho}{\partial t} + \frac{\partial}{\partial x_j}[\rho u_j] = 0 \qquad (1)$$

$$\frac{\partial}{\partial t}(\rho u_i) + \frac{\partial}{\partial x_j}[\rho u_i u_j + p\delta_{ij} - \tau_{ij}] = 0, \quad i = 1,2 \qquad (2)$$

At higher solid concentrations when the two-way coupling needs to be included, the governing equations are,

$$\frac{\partial(\alpha_f \rho_f)}{\partial t} + \nabla.\left(\alpha_f \rho_f \vec{u}_f\right) = S_{mass} \qquad (3)$$

where $\alpha_f$ is the volume fraction of the fluid phase which is the primary phase in our domain. Here $S_{mass}$ is the source term describing mass transfer between the phases.

$$\frac{\partial(\alpha_f \rho_f \vec{u}_f)}{\partial t} + \nabla.\left(\alpha_f \rho_f \vec{u}_f \vec{u}_f\right) = -\alpha_f \nabla p_f - \nabla.\left(\alpha_f \vec{\tau}_f\right) - S_p + \alpha_f \rho_f \vec{g} \qquad (4)$$

where the source term $S_P$ is given as

$$S_p = \frac{1}{V_{cell}} \int_{V_{cell}} \sum_{i=0}^{N_p} V_i \frac{\beta}{\alpha_p} (\vec{u}_f - \vec{u}_p) \delta(x - x_p) dV \qquad (5)$$

where $\beta$ is the drag force coefficient and $\alpha_p$ is the volume fraction of the solid phase (secondary phase).

The source term given by Eqn. (5) is only active at the center of the particle. The motion of every individual particle with mass $m_p$ and velocity $u_p$ in the system is calculated from Newton's second law. That is,

$$m_p \frac{du_p}{dt} = F_{drag} + F_{pressure} + F_{virtualmass} + F_{gravity} + F_{lift} \qquad (6)$$

### 3. Results and discussion

As mentioned before, the computational domain reflects the region in our experimental setup which is monitored and imaged by using a camera (top view). The primary phase velocity was found by the exact solution of the Couette flow domain. The gravity, lift and virtual mass are negligible in this two-dimensional analysis. Therefore, Eqn. (6) can be rewritten as follows:

$$\frac{du_p}{dt} = C_{drag}(u_f - u_p) \qquad (7)$$

where $C_{drag}(u_f - u_p)$ represents the drag force per unit mass and $C_{drag}$ is given as

$$C_{drag} = \frac{18\mu}{\rho_p d_p^2} \frac{C_d Re}{24} \qquad (8).$$

$u_p$ is the solid particle velocity, $\rho_p$ and $d_p$ are the density and diameter of the spherical particle. The Reynolds number is calculated based on the relative velocity of the primary (fluid) and secondary phase (solid) as

$$Re = \frac{\rho d_p |u_f - u_p|}{\mu} \qquad (9)$$

The particle velocity and displacement were calculated by integrating the two-dimensional governing equation of motion. These simulations were performed by the accuracy of the second order. In all numerical cases $dt < 10^{-6}$ sec and $dx < 10^{-4} m$.

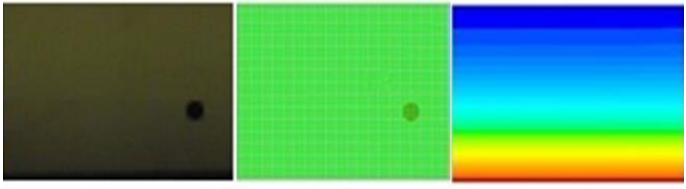

**Fig. 2** Left: particle located in the experimental domain; Middle: particle located in the meshed primary phase; Right: fluid phase under the moving boundary condition from bottom

In Fig. 2, the left photograph shows the top view of the experimental arrangement of the particle suspended in the stratified flow. The middle image is from the two-dimensional CFD domain of the discretized fluid domain with the solid particle which reflects the experimental particle location in the fluid. The right image shows the velocity contour plot of the fluid in the domain with the moving boundary condition at the bottom. The top boundary condition is zero velocity due to the symmetry of the domain on the mid-axis. The moving bottom boundary condition reflects the moving belt in the experimental setup. Since the whole domain in the experimental set up consists of two belts which move in different directions with the same velocity, just half of the domain is considered in the modeling and simulation study.

A constant velocity boundary condition is adopted from the experimental measurements of the belt and is used in the numerical calculations. The dimensionless relative velocity of the particle with respect to the moving belt was defined as

$$\emptyset = \frac{u_{belt} - u_p}{u_{belt}} \quad (10)$$

where $u_{belt}$ is the average belt velocity of the experimental setup. This average velocity of the belt was used as the boundary condition in the CFD analysis as well.

The dimensionless relative velocity of the particle, $\emptyset$, was evaluated and the results versus a range of low Reynolds number are shown in Fig. 3. As the Reynolds number increases, the error value increases; implying more uncertainties and mismatch between the experimental and numerical results. This can be attributed to several reasons including: increasing the effect of belt slipping, increasing the contribution of rotation of particles and their roughness. Here, the experimental data were calibrated based on the slippery rate of the belts over pulleys. In the experimental domain, the stratified primary phase consists of layers of fluid components with different densities while in the simulation an average value of both fluid layers was used. Therefore, the mismatch between the experimental and numerical results is expected. It should be noted that the uncertainties were quantified in both experimental and computational studies. The simulation results are about 2% sensitive to the initial location of the particle which was measured by image processing which depends on the resolution of the images. Also, 5% error was observed due to the experimental measurement of belt slippage.

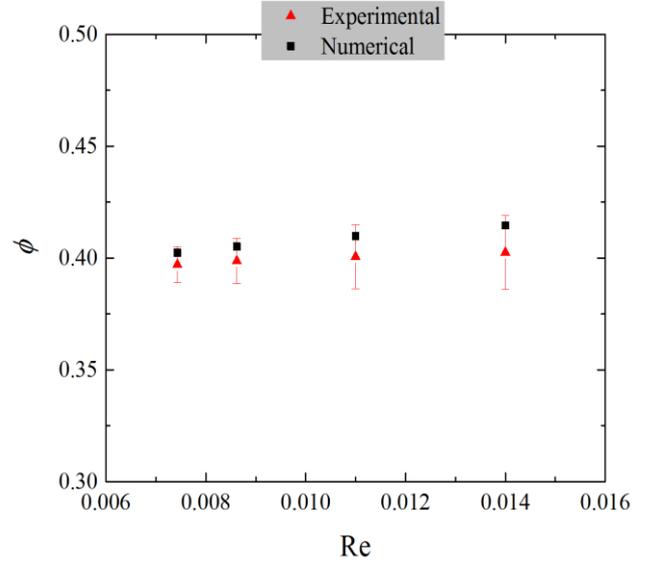

**Fig. 3** Relative dimensionless velocity versus Reynolds number

## 4. Conclusions

A computational-experimental analysis was performed on a spherical particle in a low-Reynolds-number stratified shear regime. An acceptable agreement was observed between the numerical results and the experimental data. Better agreement was observed at lower Reynolds number values. At higher Reynolds numbers the difference between the computational results and experimental data increases. One main reason for the mismatch in the trend of results may be attributed to the stratified fluid condition in the experimental apparatus which can be addressed in a three-dimensional fluid domain properly.

**References**

[1] L. G. Leal, "Particle Motions in a Viscous Fluid," Annual Review of Fluid Mechanics, vol. 12, pp. 435-476, 1980.
[2] Marc S. Ingber, "Combined static and hydrodynamic interactions of two rough spheres in nonlinear shear flow," Journal of Rheology, vol. 54, no. 4, 2010.
[3] Marc S. Ingber and Alexander Zinchenko, "Semi-analytic solution of the motion of two spheres in arbitrary shear flow," International Journal of Multiphase Flow, vol. 42, pp. 152-163, 2012.
[4] N. Fathi, M. Ingber, and P. Vorobieff, "Particle interaction in oscillatory Couette and Poiseuille flows," Bulletin of the American Physical Society, vol. 58, 2013.

[5] N. Fathi, M. Ingber, and P. Vorobieff, "Particle behavior in linear shear flow: an experimental and numerical study," Bulletin of the American Physical Society, vol. 57, 2012.

[6] N. Fathi and P. Vorobieff, "Spherical Particles in a Low Reynolds Number Flow: A V&V Exercise," ASME Verification and Validation Symposium, 2013.

[7] S. Mortazavi and G. Tryggvason, "A numerical study of the motion of drops in poiseuille flow. part 1. lateral migration of one drop," Journal of Fluid Mechanics, vol. 411, pp. 325-350, 2000.

[8] M. Razi and M. Pourghasemi, "Direct Numerical Simulation of deformable droplets motion with uncertain physical properties in macro and micro channels," Computer & Fluids, vol. 154, no. 1, pp. 200-210, 2017.

[9] X. Chen et al., "Inertial migration of deformable droplets in a microchannel," Physics of Fluids, vol. 26, no. 11, 2014.

[10] Y. Nakayama, K. Kim, R. Yamamoto, "Simulating (electro) hydrodynamic effects in colloidal dispersions: Smoothed profile method". The European Physical Journal E, 26(4), 2008, 364-368.

[11] Y. Nakayama, R. Yamamoto, "Simulation method to resolve hydrodynamic interactions in colloidal dispersion." Physical Review E, 71, 2006, 036707-1-036707-7.

[12] Popova, Marina, Peter Vorobieff, Marc S. Ingber, and Alan L. Graham. "Interaction of two particles in a shear flow." Physical Review E 75, no. 6 (2007): 066309.

[13] Ingber, Marc, and Peter Vorobieff. Localized Scale Coupling and New Educational Paradigms in Multiscale Mathematics and Science. No. DOE-UNM-25705. The University of New Mexico, 2014.

[14] Fluent, A., 2017. 18.0 ANSYS Fluent Theory Guide 18.0. Ansys Inc.